# Response of Timepix detector with GaAs:Cr and Si sensor to heavy ions

S.M.Abu Al Azm[a], G.Chelkov[a], D.Kozhevnikov[a], A.Guskov[a], A. Lapkin[a], A. Leyva Fabelo[a,c], P.Smolyanskiy[a,b,*], A.Zhemchugov[a]

[a]Joint Institute for Nuclear Research, Dubna, Russia

[b]Saratov State University, Saratov, Russia

[c]CEADEN, Havana, Cuba

[*]Corresponding author: smolyanskiy@jinr.ru

**ABSTRACT**

The response of the Timepix detector to neon ions with kinetic energy 77 and 158.4 MeV has been studied at the cyclotron U-400M of the JINR Flerov Laboratory of Nuclear Reaction. Sensors produced from gallium arsenide compensated by chromium and from silicon are used for these measurements. While in Timepix detector with Si sensor the well-known so-called "volcano effect" observed, in Timepix detector with GaAs:Cr sensor such effect was completely absent.

In the work the behavior of the Timepix detector with GaAs:Cr sensor under irradiation with heavy ions is described in comparison with the detector based on Si sensor. Also the possible reason for absence of "volcano" effect in GaAs:Cr detector is proposed.

## 1. INTRODUCTION

The manifest interest in the introduction of the advanced chromium compensated gallium arsenide (GaAs:Cr) semiconductor, developed at the Tomsk State University [1], for radiation detection continuously grows. This is a consequence of its numerous advantages [1-3] converting GaAs:Cr into a very promising material for the development of sensors for applications ranging from medical imaging to high-energy physics.

The JINR Dzhelepov Laboratory of Nuclear Problems in cooperation with other institutions such as Medipix international collaboration (CERN), Institute of Experimental and Applied Physics of Czech Technical University (Prague) and the Tomsk State University investigates the convenience of using the GaAs:Cr as integrated sensor in hybrid pixel detectors based on the Timepix readout chip [4].

Obtained to date results on the detection of photons using the Timepix detector with GaAs:Cr sensor are very promising, and besides being published [5], a prominent practical result was reached when the implementation of this hybrid pixel detector in a fully operational CT scanner was achieved [6].

The investigations continue today and the response of the device to other types of radiation such as heavy ion beams is evaluated. In this research process a useful complementary method is the comparison between the results obtained by the Timepix detector based on GaAs:Cr sensor with those obtained by a similar detector but made using a silicon sensor. This procedure allows testing the behavior of the former comparing it with the considered as classic silicon sensor. Also let studying the characteristics and behavior of the Timepix with Si sensor, that is still in development and improvement.

The main goal of this work is the investigation of the response of Timepix detectors with GaAs:Cr to heavy ions. While Timepix detector with Si sensor was irradiated by heavy ions by several research groups [7,8], there were no any information about similar results with GaAs:Cr Timepix detector so far, therefore this objective seems an interesting problem.

## 2. MATERIALS AND METHODS

The Timepix detector is a hybrid detector consists of a pixelated semiconductor sensor bump-bonded to a readout electronics with matrix of 256×256 independent channels (pixels) with pitch 55 μm. Each channel with respective preamplifier, discriminator and digital counter can independently work in one of three modes: Medipix mode (the counter counts number of detected particles), Timepix mode (the counter measures the time between the moment the particle is detected and the end of the frame) and Time-over-Threshold (TOT) mode (the counter is used as a Wilkinson type ADC allowing direct energy measurement in each pixel).

In this work we used two Timepix detectors based on Si and GaAs:Cr sensor and operated in TOT mode. Table 1 shows main characteristics of these detectors. Whereas the Timepix contains 65536 independent channels and their response can never be identical it is necessary to perform an energy calibration for each pixel. Such calibration was done with characteristic X-ray radiation [9].

*Table 1. Main detector characteristics and the used biased voltage intervals*

| Sensor material | Sensor thickness [µm] | Readout chip | USB Interface | $U_{bias}$ [V] |
|---|---|---|---|---|
| Si | 300 | Timepix DO5 | FitPix | from +5 to +100 |
| GaAs:Cr | 300 | Timepix SO2 | FitPix | from -5 to -200 |

The irradiation process took place in one of the output channels cyclotron U400M of the JINR Flerov Laboratory of Nuclear Reaction where $^{22}$Ne ions were accelerated to energies 77 and 158.4 MeV. The detectors were irradiated simultaneously placing side by side in the vacuum chamber as shown at figures 1 and 2. The distance between the centers of both sensitive detector areas was 3 cm. The temperature conditions in the chamber were about 27 $^0$C. The ion beam cross section was larger than the area occupied by both sensors, and sufficiently homogeneous to consider that the detectors were under the same irradiation conditions.

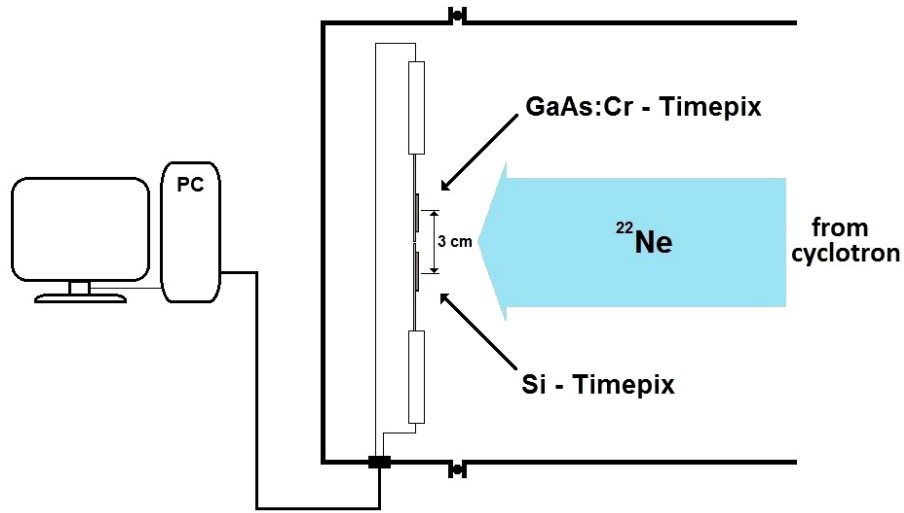

*Figure 1: General schema of the experiment showing the location of the two detectors relative to the incident ion beam inside the irradiation chamber*

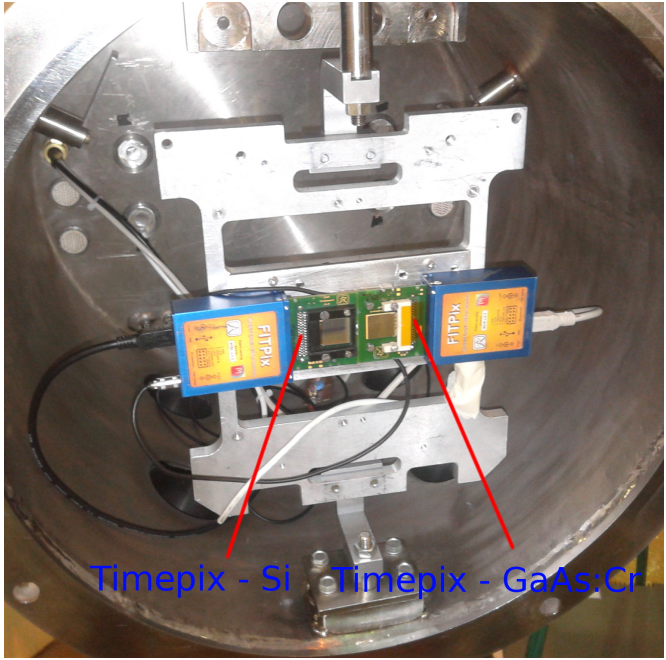 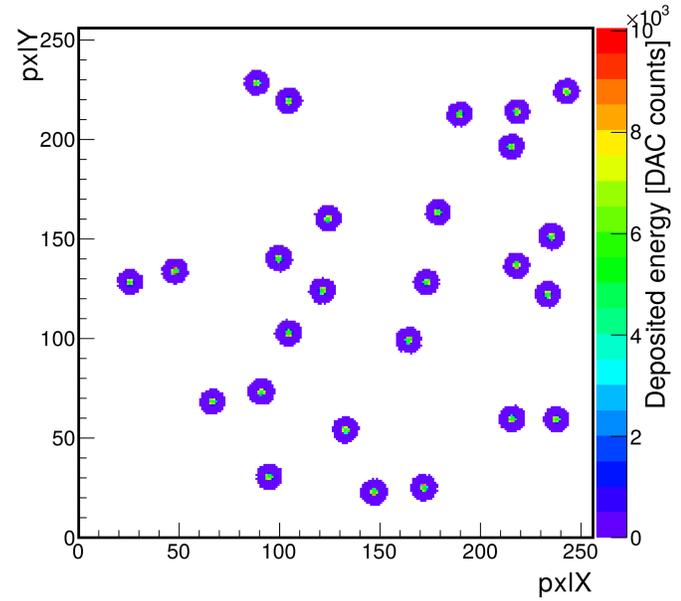

*Figure 2: Photo of the vacuum chamber where are visible the two Timepix detectors with readout electronics FitPix*

*Figure 3: Typical frame recorded by Timepix detector with clusters created by heavy ions*

The energy deposited by a charged particle in the semiconductor sensor results in an induced charge in adjacent pixels the signals of which form a cluster (fig. 3). The full deposited energy by the particle is defined as the sum of the energies of all pixels in the cluster. Due to a spatial resolution and single event counting possibility of the Timepix each individual ion can be observed together with clusters from other particles presented in the irradiation chamber. A correlation between specific types of ionizing radiation and the obtained cluster shapes as well as the possibility to differentiate different types of irradiation by the pattern they introduce in Medipix2 detectors were described in [10,11]. With this capability in our analysis for obtaining the best results we selected only clusters produced by single ion and excluded the clusters formed by pile-up of several clusters from ions or clusters from other particles presented in the beam.

Recording the data from both detectors was done with USB interface FitPix [12] and the Pixelman software package [13]. All data processing was carried out in the ROOT data analysis framework [14].

## 3. RESULTS AND DISCUSSION

### 3.1 The response of the Timepix detector with Si sensor

Using the code system SRIM-2013 [13] the range of $^{22}Ne$ ion in silicon for the two energies was calculated, resulting in 41.8 µm for 77 MeV, and 112 µm for 158.4 MeV. This confirms that the silicon sensor with a thickness of 300 microns is able to stop 100% of the ions to the come to him, i.e. all the incident ions will interact and deposit all their energy into the sensor. More than 99.7% of the deposited energy is lost in processes associated with the ionization, as can be seen in Table 2.

*Table 2. The ion ranges and energy losses of $^{22}Ne$ ion interacting with the sensor materials for the two studied energies*

|  | Si (Ne, 77 MeV) | Si (Ne, 158.4 MeV) | GaAs:Cr (Ne, 77 MeV) | GaAs:Cr (Ne, 158.4 MeV) |
|---|---|---|---|---|
| Longitudinal ion range [µm] | 41.8 | 112 | 26.3 | 67.1 |
| Radial ion range [µm] | 0.93 | 1.58 | 1.06 | 1.66 |
| Total ionization energy loss [keV/Ion] | 76829.2 (99.77 %) | 157806.0 (99.877 %) | 76759.7 (99,688 %) | 157728.3 (99,828 %) |
| Total phonon energy loss [keV/Ion] | 160.1 (0,20 %) | 182.0 (0,115 %) | 229.3 (0,298 %) | 259.4 (0,164 %) |
| Total target damage energy loss [keV/Ion] | 10.72 (0,01 %) | 12.04 (0,008 %) | 10.99 (0,014 %) | 12.28 (0,008 %) |

Figures 4 and 5 show the energy spectra and the average cluster profiles respectively, for different bias voltage values obtained for different values of bias voltage and two incident ion energies. The average cluster profile corresponds to profile of generalized cluster (10000 typical clusters are superposed on each other in their geometrical center and normalized).

Analyzing the response of the detector to 77 MeV ions can be seen that, for the bias voltage +5V the mean energy in cluster is about 62 MeV. This value is lower than the expected 77 MeV, which, according to the simulation, should have been deposited in the sensor. At this bias voltage this is related with the incomplete charge collection and incomplete depletion volume of silicon sensor [14] due the insufficient electric field applied to the sensor work volume. However, in this case the average cluster profile has a typical Gaussian-like shape.

When the bias voltage is increasing to +15 and +20 V the responses are characterized by the high values of mean energy in cluster, exceeding 77 MeV, probably as a consequence of the nonlinearity in used energy calibration at high energies [7,17]. But it is interesting that even though both profiles increased its height, in +15 V a deformation starts to appear in the shape of the peak top. For + 20V this deformation has become in an evident unusual depression, that called "volcano effect" because of its profile [7,8].

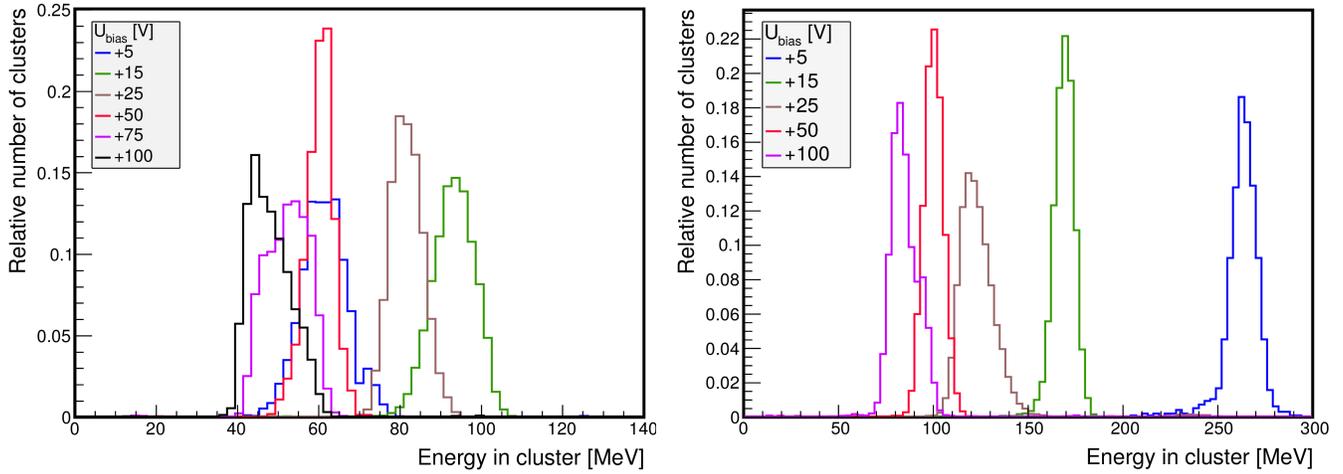

*Figure 4. Energy spectra registered by Timepix detector with Si sensor for different bias voltages and ion energy 77 MeV (left) and 158.4 MeV (right)*

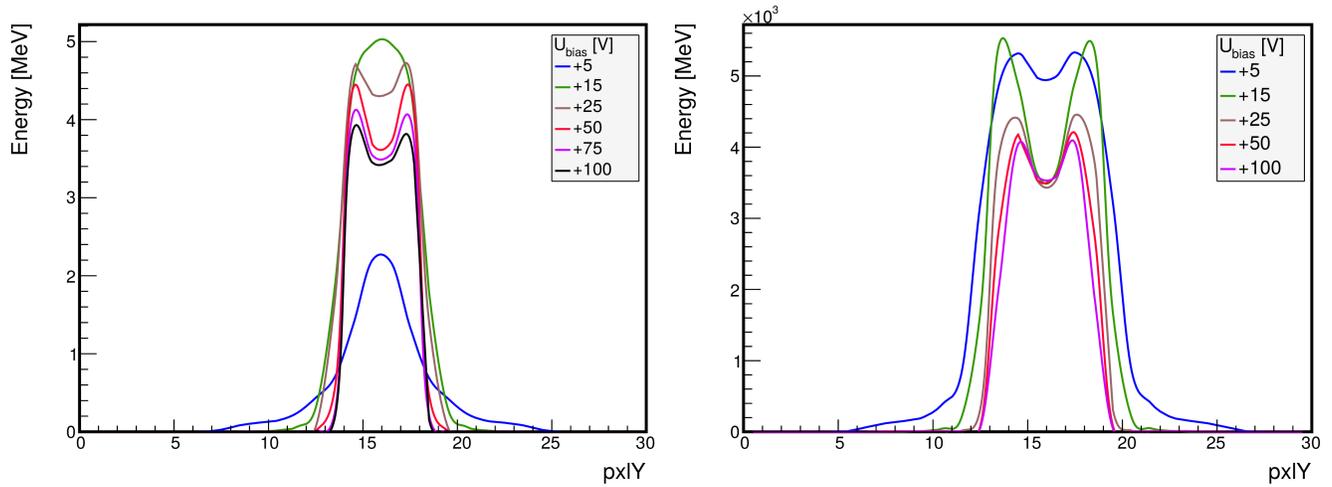

*Figure 5. Average cluster profiles registered by Timepix detector with Si sensor for different bias voltages and ion energy 77 MeV (left) and 158.4 MeV (right)*

For higher bias voltages in the mean energy cluster is always less than 77 MeV, but the "volcano effect" is enhanced significantly.

Analyzing the detector response to 158.4 MeV $^{22}$Ne ions (see the right images in Figures 4 and 5) is observed that the previously behaviors are repeated, but now the "volcano effect" appears from the lowest biasing values.

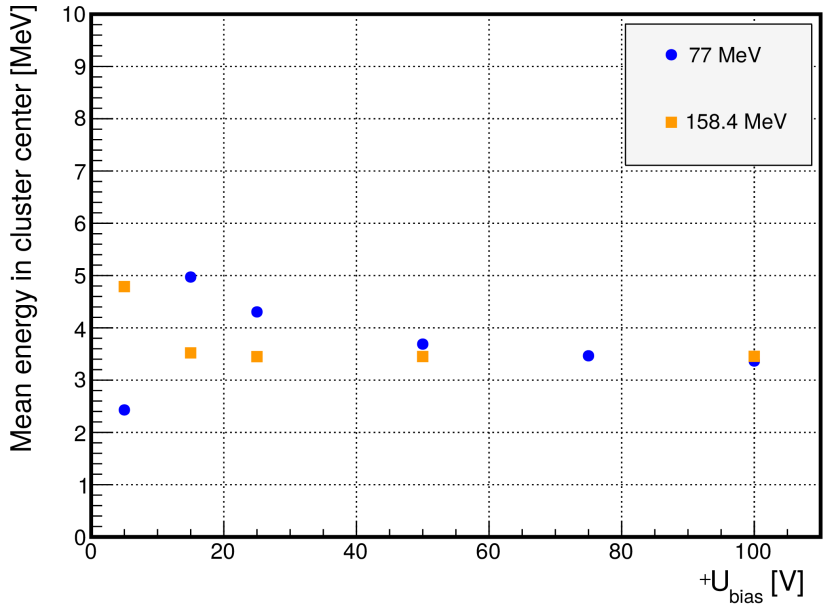

*Figure 6. Dependence of the mean energy in the geometrical center of cluster with the bias voltage for 77 and 158.4 MeV ion energies (for Timepix detector with Si sensor)*

Figure 6 shows the dependence of the mean energy in cluster geometrical center with the bias voltages in the Timepix detector with Si sensor for the two ion energies. This supplementary information was obtained from the average cluster profiles shown in figure 5, and allows us to see clearly how with increasing bias voltage takes place a process of signal saturation to the same level for both ion energies. This behavior is consistent with the saturation of the preamplifier in each channel of the Timepix electronics.

### 3.2. The response of the Timepix detector with GaAs:Cr sensor

Whereas the response of the Timepix detector with Si sensor to heavy ions was already investigated by several groups from all over the world (see e.g. [7,8]), there were no articles about the irradiation of the Timepix with GaAs:Cr sensor with heavy ions so far.

As shown in the Table 2, the 77 and 158.4 MeV $^{22}$Ne ions also reaches deposit all its energy into the volume of the 300 µm GaAs:Cr sensor, and losses by ionization are still the dominant processes.

The energy spectra obtained using the Timepix detector with GaAs:Cr sensor for different bias voltages are shown in figure 7, while figure 8 shows the average cluster profiles also for certain values of biasing values. As one can see from figure 7 the mean energy in cluster is much lower than the expected 77 and 158.4 MeV, that indicates the used energy calibration is incorrect for heavy ion detection.

The first remarkable fact observed in these figures is that a significant reduction in the charges collection efficiency takes place for both ion energies, compared with the experiment where the silicon sensor was used. However now observed before "volcano effect" has completely disappeared.

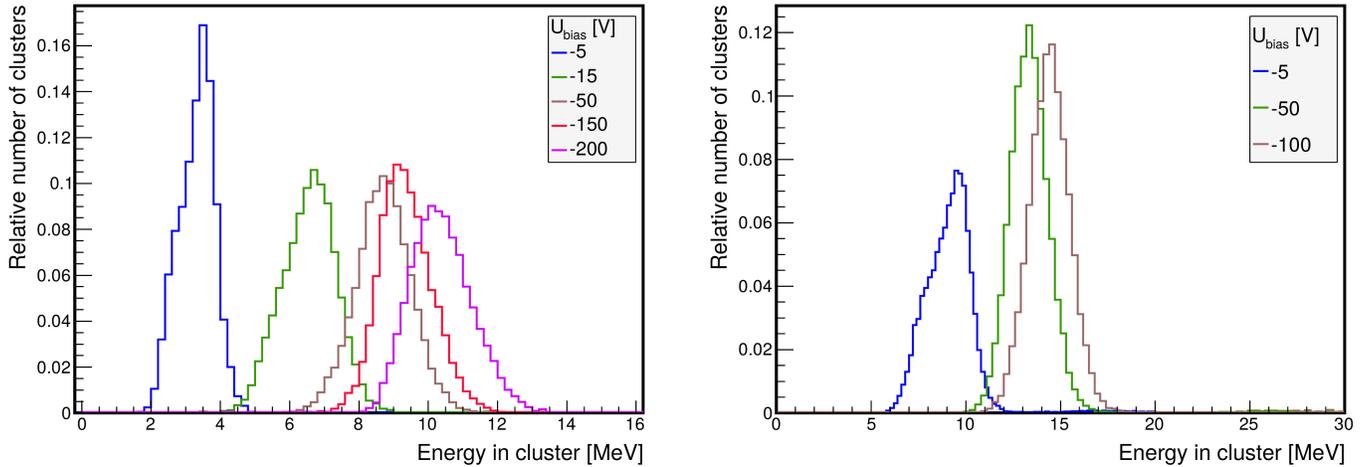

*Figure 7. Energy spectra registered by Timepix detector with GaAs:Cr sensor for different bias voltages and ion energy 77 MeV (left) and 158.4 MeV (right).*

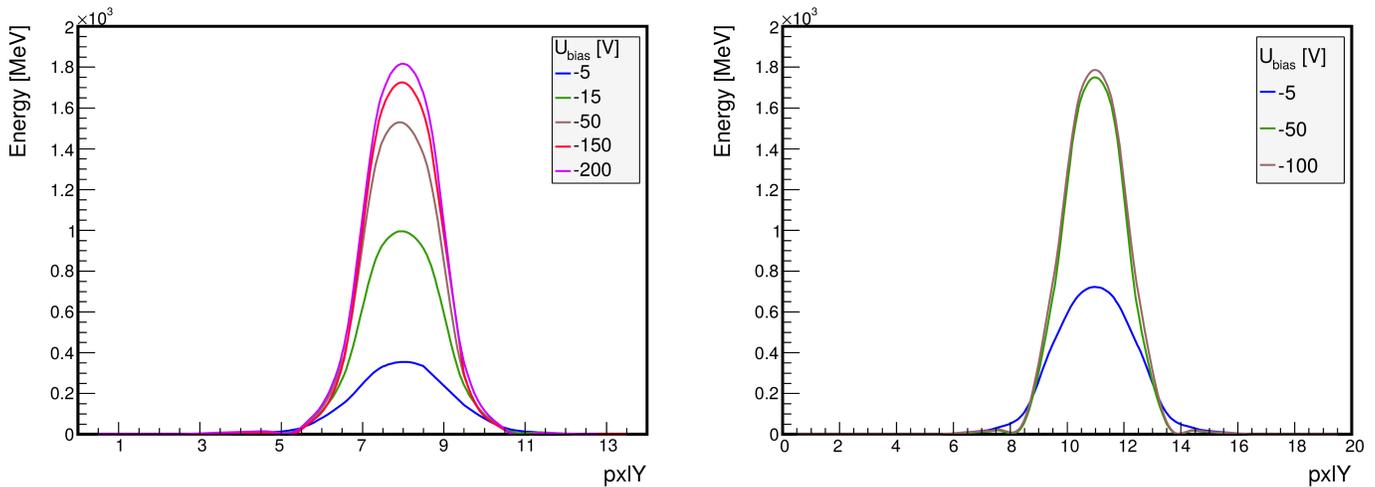

*Figure 8. Average cluster profile registered by Timepix detector with GaAs:Cr sensor for different bias voltages and ion energy 77 MeV (left) and 158.4 MeV (right).*

Figure 9 shows how the doubling the energy of ions led to increase of mean energy in geometrical cluster center twofold only for low bias voltage at which there is incomplete charge collection and consequently input charge for preamplifier is a small. At higher bias voltages, as observed before for the Si sensor, the maximum signal in cluster significantly reduces their growth to become practically imperceptible, indicating that the input charge is high enough to produce the preamplifier saturation.

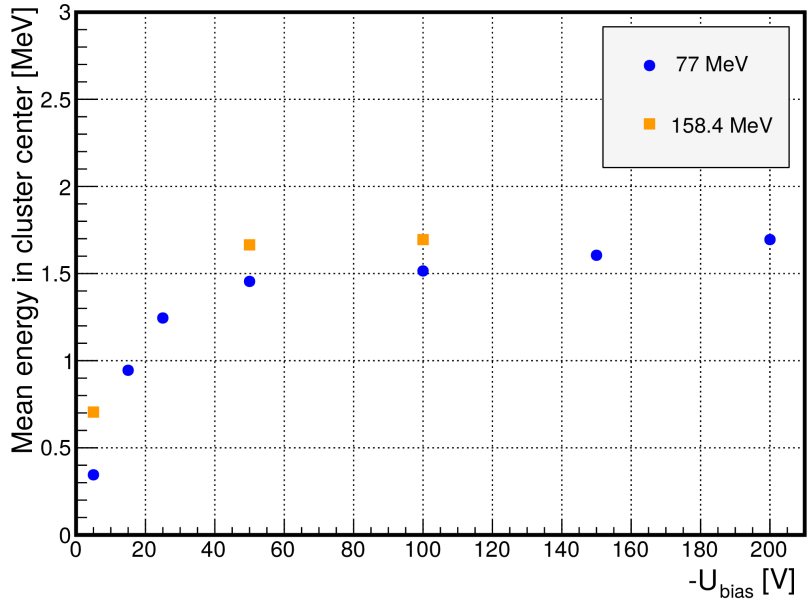

*Figure 9. Dependence of the mean energy in the cluster geometrical center with the bias voltage for 77 and 158.4 MeV ion energies.*

The manifestation of "volcano effect" is compatible with the fact that the Timepix readout chip is equipped with an internal protection circuit for very high input charges that is active only for hole collection mode (for electron collection mode internal protection circuit is not provided). $^{22}$Ne ions with kinetic energy 77 and 158.4 MeV generate very high numbers of electron-holes pairs in the sensor (Table 2) and for Timepix with np-type Si sensor it triggers the chip overload protection system to limit the portion of the signal exceeds the predetermined threshold that is expressed in the reduction of signal from pixels with high input charges (usually in the center of cluster). Thus, as was shown the "volcano effect" is present in Timepix with Si sensor, but absolutely absent in Timepix with GaAs:Cr sensor.

Finally, figure 10 shows the dependence of the mean cluster radius with the bias voltage for both studied Timepix detectors irradiated with 77 MeV $^{22}$Ne ions.

According to the results obtained from the simulation with SRIM-2013, the approximate dimensions of the primary electron-hole cloud generated by 77 MeV ions in Si are 41.8 µm long and 0.93 µm wide, whereas these same ions in the GaAs:Cr create an electron-hole cloud of 26.3 µm long and 1.06 µm wide. As noted, the two charge carrier clouds have both in silicon and in gallium approximately the same radius. From figure 10 one can observe that the Timepix clusters have in average a radius of 450 µm for Si and 250 µm for GaAs:Cr (pixel size is 55 µm). This means that the cluster radius ratio between Si and GaAs:Cr is in the range 1.5 – 2 pixels. The observed difference between this relationship and the one existing between the radius of the electron-hole clouds at the initial collection moment is the result of the multiple differences in the intrinsic properties of the two sensors materials, decisive in the charge transport and collection processes.

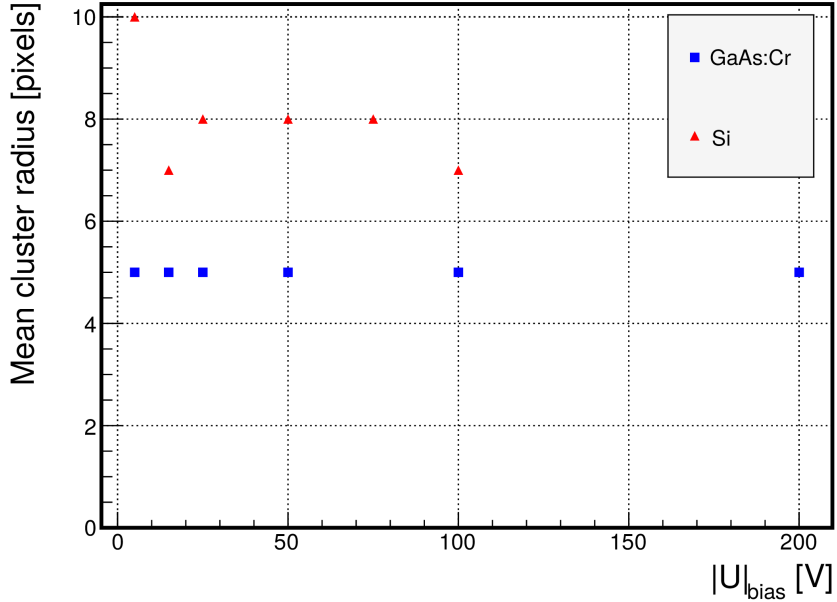

*Figure 10. Dependence of the mean cluster radius with the bias voltage for the two sensors.*

## 4. CONCLUSIONS

It was shown that the Timepix detector with GaAs:Cr sensor can register heavy ion in energy range 77- 158.4 MeV as well as Timepix with Si. The used energy calibration with characteristic X-rays is incorrect for heavy ion detection due to the pixel signal deviates at about energy 1 MeV per pixel [7,17]. During investigation of current work it was confirmed the "volcano effect" is present in Timepix with Si sensor (np-type), but absolutely absent in Timepix with GaAs:Cr sensor. The possible explanation of this fact connected with an internal protection circuit for very high input charges in hole collection mode is given.

## 5. ACKNOWLEDGEMENTS

This work is supported by the Ministry of Education and Science of Russian Federation under the contract No. 14.618.21.0001. The authors would like to express thanks the contributions of colleagues from FLNR, in particular V.A.Skuratov and S.A.Mitrofanov.